\pdfoutput=1
\documentclass[cits]{JINST}
\newcommand{\unit}[1]{\ensuremath{\mathrm{\,#1}}}
\renewcommand{\u}[1]{\unit{#1}}

\DeclareFontFamily{U}{euc}{}
\DeclareFontShape{U}{euc}{m}{n}{<-6>eurm5<6-8>eurm7<8->eurm10}{}%
\DeclareSymbolFont{AMSc}{U}{euc}{m}{n} 
\DeclareMathSymbol{\umu}{\mathord}{AMSc}{"16}

\usepackage{ulem}

\title{First test of a power-pulsed electronics system  on a GRPC detector in a 3-Tesla magnetic field  }

\author{
  L.~Caponetto$^a$, C.~Combaret$^a$, C.~de la Taille$^b$, F.~Dulucq$^b$, R.~Kieffer$^a$, I.~Laktineh$^a$\thanks{Corresponding author.}, N.~Lumb$^a$, L.~Mirabito$^a$, N.~Seguin-Moreau$^b$.\\ 
  \llap{$^a$} Universit\'e de Lyon, Universit\'e Lyon 1, CNRS/IN2P3, 
  IPNL, 4 Rue E.~Fermi, 69622 Villeurbanne Cedex, France\\
   \llap{$^b$} Laboratoire de l'Acc\'elerateur Lin\'eaire,
Centre d'Orsay, Universit\'e de Paris-Sud XI,
BP 34, B\^atiment 200, F-91898 Orsay CEDEX, France\\
 E-mail: \email{laktineh@in2p3.fr}

}

\abstract{ An important technological step towards  the  realization of  an ultra-granular  hadronic calorimeter to be used in the future International Linear Collider (ILC) experiments has been made.   A $33\times50 ~\hbox{cm}^2$  GRPC detector  equipped with a power-pulsed electronics  board offering a $1~\hbox{cm}^2$ lateral segmentation was successfully  tested  in a 3-Tesla magnet operating at the H2 beam line of the CERN SPS.  An important reduction of power consumption with no deterioration of the detector performance is obtained  when the power-pulsing mode is applied.  This important result shows that ultra-granular calorimeters for ILC experiments are not only an attractive but also a realistic option. }
\begin{document}


\section{Introduction}

One of the attractive techniques that the future ILC experiments can use to achieve unprecedented jet energy measurement resolution is the one based on the Particle Flow Algorithms  (PFA) approach~\cite{PFA}.  In this approach, particles are tracked in the different sub-detectors and their energy or momentum is estimated in the most appropriate sub-detector according to their nature. This requires electromagnetic and hadronic calorimeters to have a tracking capability in addition to their usual functionality.  To achieve this a new generation of sampling hadronic calorimeters with high granularity in both the transverse and the longitudinal directions is being developed by the CALICE collaboration~\cite{CALICE}. One of those calorimeters which is proposed as an HCAL option in the Letter Of Intent (LOI) submitted by the  ILD collaboration~\cite{ILD_LOI} is a hadronic calorimeter  with GRPCs as a sensitive medium offering a  $1~\hbox{cm}^2$ lateral segmentation.  To cope with the more than  50 million electronic channels needed  for such a high-granularity but still compact and hermetic calorimeter, a  new electronic chip (ASIC) called HARDROC~\cite{HR}  was developed. The chip, which was conceived to be very thin in order to be embedded into the detector  to ensure the needed  compactness, features a very  low power consumption (<1\u{mW/channel}) when operated continuously.  Even with such low consumption it is almost impossible to envisage the use of such calorimeters without a very sophisticated cooling system which increases the complexity of the detector design and a loss of its compacity.%
To overcome this difficulty, one can take advantage of the duty cycle of the future ILC: one millisecond bunch crossing every 200\u{ms}. The HARDROC chip was therefore equipped with a new mechanism called power-pulsing scheme (PP hereafter). This scheme allows to keep the chip idle during the inter-bunches  and to power it just before the bunch crossings.  This leads to an important reduction factor of the chip power consumption (>100) when applied within the ILC duty cycle scenario. The PP scheme was first successfully tested on a test board equipped with just one chip. It was then tested on an electronics board hosting 24 chips connected to each other using a daisy chain system.  In order to validate the PP scheme in the future ILC conditions and more particularly in the presence of  a strong magnetic field (> 3\u{Tesla}) the 24-ASIC board was mounted on a GRPC detector similar to those proposed for the semi-digital hadronic calorimeter in the ILD LOI and tested in a magnetic field of 3\u{Tesla} on the H2 beam line of the CERN SPS using a pion beam.\\
In this paper the HARDROC chip will be introduced. The power-pulsing scheme and the results obtained using the single-chip test board will be described. Afterwards, the setup of the GRPC detector and the associated electronics will be presented  together  with the acquisition system used in the 3-Tesla test beam to demonstrate the validity of the PP scheme. Finally the results obtained during the test beam campaign will be briefly discussed.

\section{Power-pulsed electronics }

A detailed description of the HARDROC ASIC used to read out  the GRPC signal can be found in reference~\cite{HR}. Only a short description of the ASIC, emphasizing its power-pulsing feature, will be given hereafter.%

The HARDROC chip has  64 input channels, each using three comparators to digitize the signal. To set the comparator's reference levels (thresholds), three 10-bit DACs are embedded into the ASIC (thresholds settings in the 0-1023 DAC units range). For each of these three comparators, a logic OR is made of the 64 channels. Then, the status of the selected OR is evaluated by the common digital part.  %
If one of the 64 comparators is fired, the data are stored in the memory. This is  known as the "auto-triggering" scheme.  An important feature of the ASIC is the possibility to adjust the gain of each of the 64 channels by a factor between 0 and 2 with a 8-bit precision.  The ASIC can store up to 127 frames in its internal memory.  
A frame stored in one ASIC consists of a hit map of the 64 channels with 2 bits per channel plus a time stamp of 24 bits and an 8-bit chip identifier. The Gray-coded time stamp is derived by a 5\u{MHz} clock.  Each channel of the ASIC has a test capacitor of 2$\pm$0.02\u{pF} which can be used to calibrate its response. %
This is a useful tool to make the response of the different channels as uniform as possible~\cite{small}. %
The cross-talk between two channels of one ASIC was measured by injecting an electric signal equivalent to a MIP ($\approx$1\u{pC}) through one channel. %
The signal observed in the other channels was found to be less than 2\u{\%} of the injected charge.%

The communication between the acquisition system (DAQ) and the ASIC acts in two steps: an acquisition phase and a readout phase.  The acquisition phase is stopped at the end of the acquisition window or when a full-memory state is reached. Then, triggered by an external signal provided by the DAQ, the second phase starts:  after receiving the StartReadout signal, each ASIC enters its readout phase  and issues an EndReadout upon completion. The latter is transmitted to the next ASIC which interprets it as a StartReadout command. %
To read out large GRPC detectors with a $1\,\hbox{cm}^2$ lateral resolution many ASICs need to be assembled together.  To limit the number of wires between the  ASICs and the DAQ system the digital readout signals are connected with an open collector bus per board  and daisy-chained in a ring topology, leading to use only one wire for the link. Similarly, a daisy-chained ring bus is used for the assignment of  configuration parameters (called Slow Control hereafter) to all the ASICs of one board. %
The situation is different for the acquisition phase where all the ASICs should start at the same time.  In this case, a StartAcquisition command is broadcasted to all the ASICs to initiate this phase. 
  All the commands sent to the ASICs by the DAQ are controlled by  an FPGA device. %

\begin{figure}[ht]
      \begin{center}
        \includegraphics*[width=.5\textwidth,keepaspectratio]{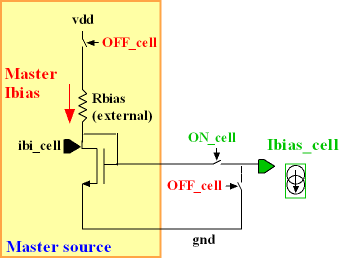}
	\caption{Schematic view of the power-pulsing mode \label{PP1}}
     \end{center}

\end{figure}

Since the ILC duty cycle is expected to be one millisecond bunch crossing every 200\u{ms}, the HARDROC was conceived to take advantage of such a scenario using the power pulsing mode thanks to a special module called Power-On-Digital (POD).  This consists of switching off the internal bias currents of the various ASIC's channels during the inter-bunches whereas the power supplies are kept on (see Figure \ref{PP1}). There are 3 Power-On signals:      Power-On-Analog (controlling the analog part), Power-On-Digital (controlling the clock-gating) and Power-On-DAC (controlling the setting of the discriminators thresholds).  Each of them 
 can be forced  through the parameters of the Slow Control configurations. %
  The power consumption of the  Power-On-Analog, the  Power-On-DAC and the Power-On-Digital  parts has been optimized and is summarized in Table\ref{tab.consom}.

\begin{table}[h]
\begin{center}
\caption{Power consumption of the different Power-On parts per channel.\label{tab.consom}}
\begin{tabular}{|l|c|}
\hline
Power-On part& consumption ($\mu$W)  \\
\hline
Power-On-Analog   	&   1325\\
Power-On-DAC  	&   50 \\
Power-On-Digital   	&  50\\
\hline
Power-On-All   	&  1425\\
\hline
Power-On-All at 0.5\% duty cycle & < 7.5 \\
\hline
\end{tabular}
\end{center}
\end{table}

\begin{figure}[ht]
      \begin{center}
        \includegraphics*[width=.5\textwidth,keepaspectratio]{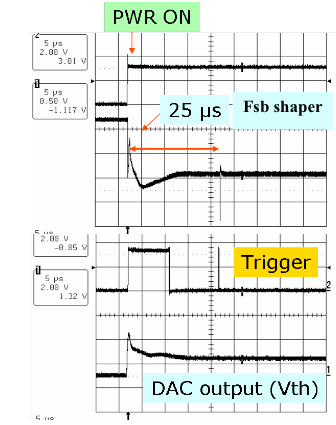}
	\caption{A screen shot showing the awake  time needed for both the Analog and the DAC  Power-On to be operational \label{PP2}}
      \end{center}

\end{figure}

The PP scheme described above was tested on one ASIC using a dedicated test board without any decoupling capacitors on the bias voltages.  The power consumption has been measured and found to be about 1.5\u{mW/channel} without power pulsing.  An ILC-like duty cycle (1ms duration and 200 ms period) has been used to control the Power-On pins as well as an input pulse synchronized with it. 
The power consumption was measured in this case and found to be 7.5$\mu$W/channel.  Another important result of this test concerns the measurement of the awake time. This is the time needed for the ASIC to be operational after a Power-On signal.  This awake time has been measured on the analog part and on the DAC part. It  was found to take 2$\mu$s for the analog part to be operational and to provide a discriminator output, and 25$\mu$s for the DAC part (see Figure \ref{PP2}) to reach its nominal value within a few mV. The DAC is slower to settle as it is filtered internally to minimize the noise and the inter-channel coupling.

\begin{figure}[ht]
      \begin{center}
        \includegraphics*[width=.8\textwidth,keepaspectratio]{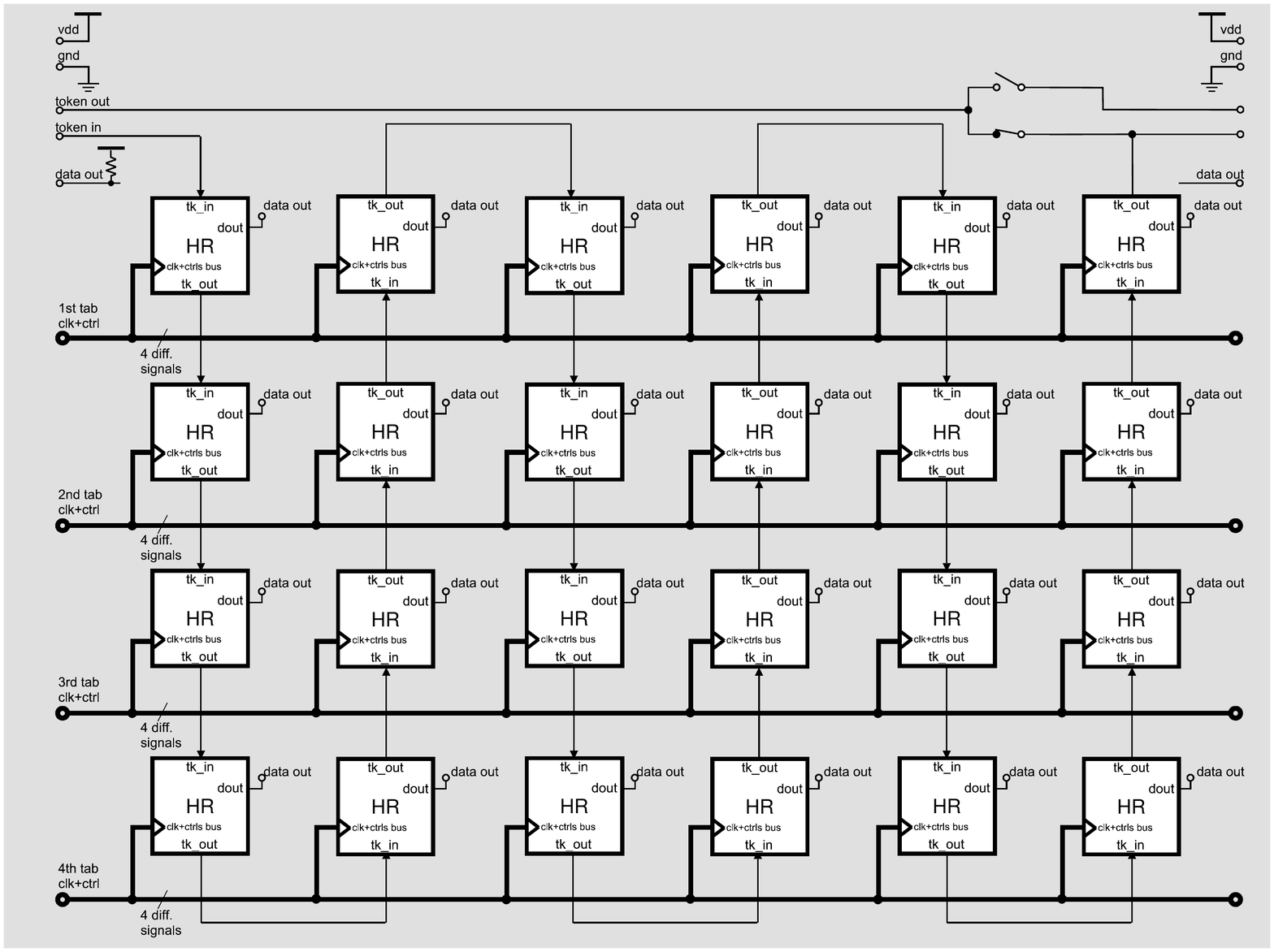}
	\caption{A schematic view of the 24-ASIC board.  \label{ASU}}
      \end{center}
\end{figure}

  The same test was realized with an electronics board conceived to host 24 HARDROCs.  The ASICs are arranged as a $6\times4$ matrix on a $50 \times 33~\hbox{cm}^2$ PCB. The board, called Active Sensor Unit (ASU), is an 8-layer  $1.2~\hbox{mm}$ thick Printed Circuit Board (PCB) which allows interfacing the $64\times24$ input channels with the GRPC detector through $1~\hbox{cm}^2$ pads printed on the face of the PCB put into contact with the detector. As shown in Figure \ref{ASU}, the 24 HARDROCs are connected in a linear topology, exploiting the HARDROC daisy chain feature.  Indeed, the logical token passing mechanism uses the two already cited signals (StartReadout and EndReadout) to ensure that only one of the HARDROCs is sending the stored digital data at a given time window. That scheme permits the DOUT (data out) open collector output pins of each HARDROC to be connected together, granting a single line data out for all the ASICs.  
  The board is connected to an acquisition board called Detector InterFace (DIF) which hosts an FPGA. The DIF  transmits the configuration parameters to the different ASICs and synchronizes the acquisition phases described before. In  addition the DIF is used to power the whole board with its 24 ASICs.%
  
 In the case of the 24-ASIC board, both the board and the DIF remain powered but as  in the case of the test bench the biases of  the ASICs are switched off during the interbunch. The DIF is configured to receive two external signals : a global one  called Power-On and another one called Acquisition-Enable. The Power-On signal starts the three parts of the  ASIC's Power-Ons:  Analog, Digital and DAC. Then the Acquisition-Enable signal is sent to the DIF and this initiates the acquisition phase.%
  Since the awake time observed here was also found to be around 25$\mu$s, a minimum delay of (50$\mu$s) was kept before sending the Acquisition-Enable signal. This is to ensure full stability of the chips before each acquisition. 
 No data can be recorded when Acquisition-Enable is low.%
   The power consumption of this board was measured with and without the power-pulsing mode. This was found to be in agreement with the values measured with individual ASICs. The power consumption of the electronics board outside  the ASICs  are found to be negligible.

\section{GRPC detector }

\begin{figure}[hh]
\begin{center}
\includegraphics[width=1.\textwidth, trim=0mm 4.5cm 0mm 6cm,clip=true ]{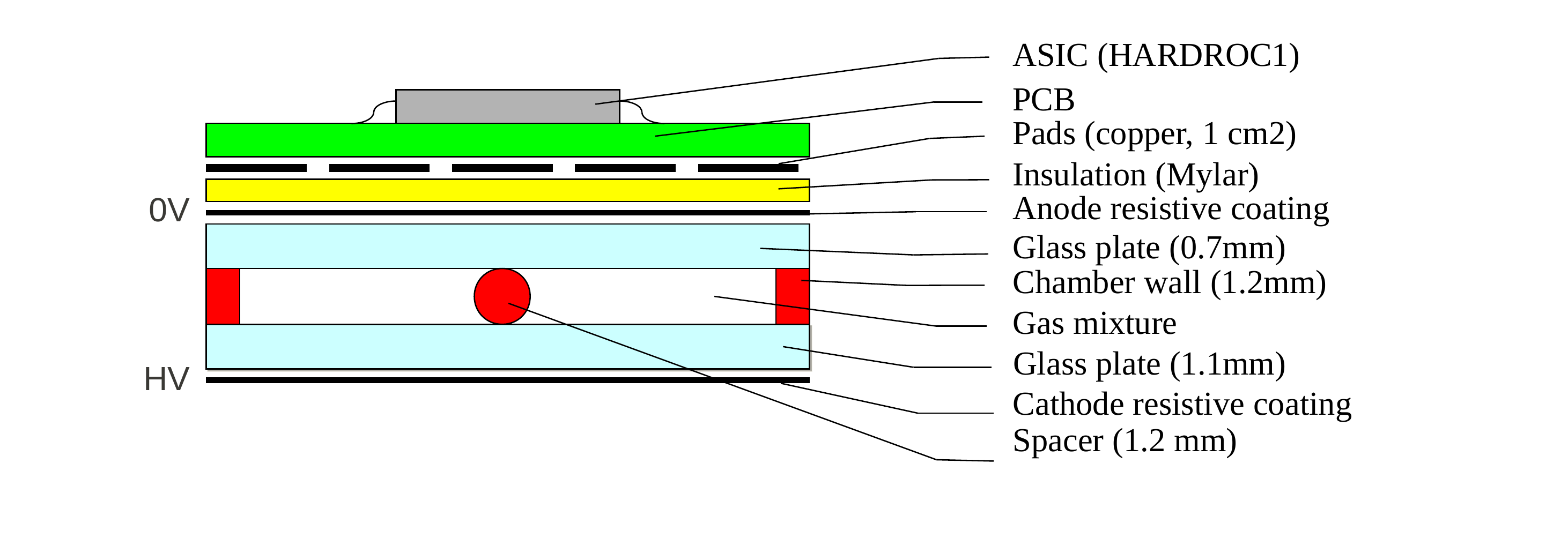}
\end{center}
\caption{Schematic view of a glass RPC\label{fig.rpc}}
\end{figure}

One of the hadronic calorimeters proposed for the future ILC experiments is the so-called semi-digital hadronic calorimeter (SDHCAL). The SDHCAL proposes to use GRPCs as a sensitive medium.  To test the PP scheme in  the future ILC experiment conditions a GRPC similar to the one proposed for the SDHCAL\cite{ILD_LOI} was built.  As shown in Figure \ref{fig.rpc},  the GRPC is made of two glass plates of the same size ($33\times50 ~\hbox{cm}^2$)  but different thickness.%
The thinner (0.7\u{mm}) glass is used to build the anode while the thicker (1.1\u{mm}) forms the cathode.   
The two plates have a bulk resistivity of $10^{12}-10^{13} \Omega$cm. They are kept separated by a 1.2\u{mm}  frame.  Ceramic  balls of  1.2\u{mm} diameter are glued on  the internal face of each plate. The balls, separated  by 10 \u{cm}, allow to maintain the same gas gap everywhere inside the chamber. %
The outer sides of the glass plates are covered by a thin layer of resistive coating and connected to a high voltage supply.  %
A 50\u{\umu m} Mylar layer separates the anode from the $1\times1\u{cm^2}$ copper pads of the electronics board.  %
The reduced anode thickness is intended to minimize the number of pads associated with the passage of one particle (multiplicity).  %
The signal is induced by the charge of the avalanche electrons.  %
The thinner the glass plate the stronger is the signal in the closest pad and the lower is the signal seen by the neighboring pads. %
A special paint with high surface resistivity ($\approx 2 \hbox{M}\Omega/\square$) was used as the resistive coating. %

\section{Experimental setup }
\begin{figure}[hh]
\begin{center}
\includegraphics[width=0.6 \textwidth, trim=0mm 4.5cm 0mm 6cm,clip=true ]{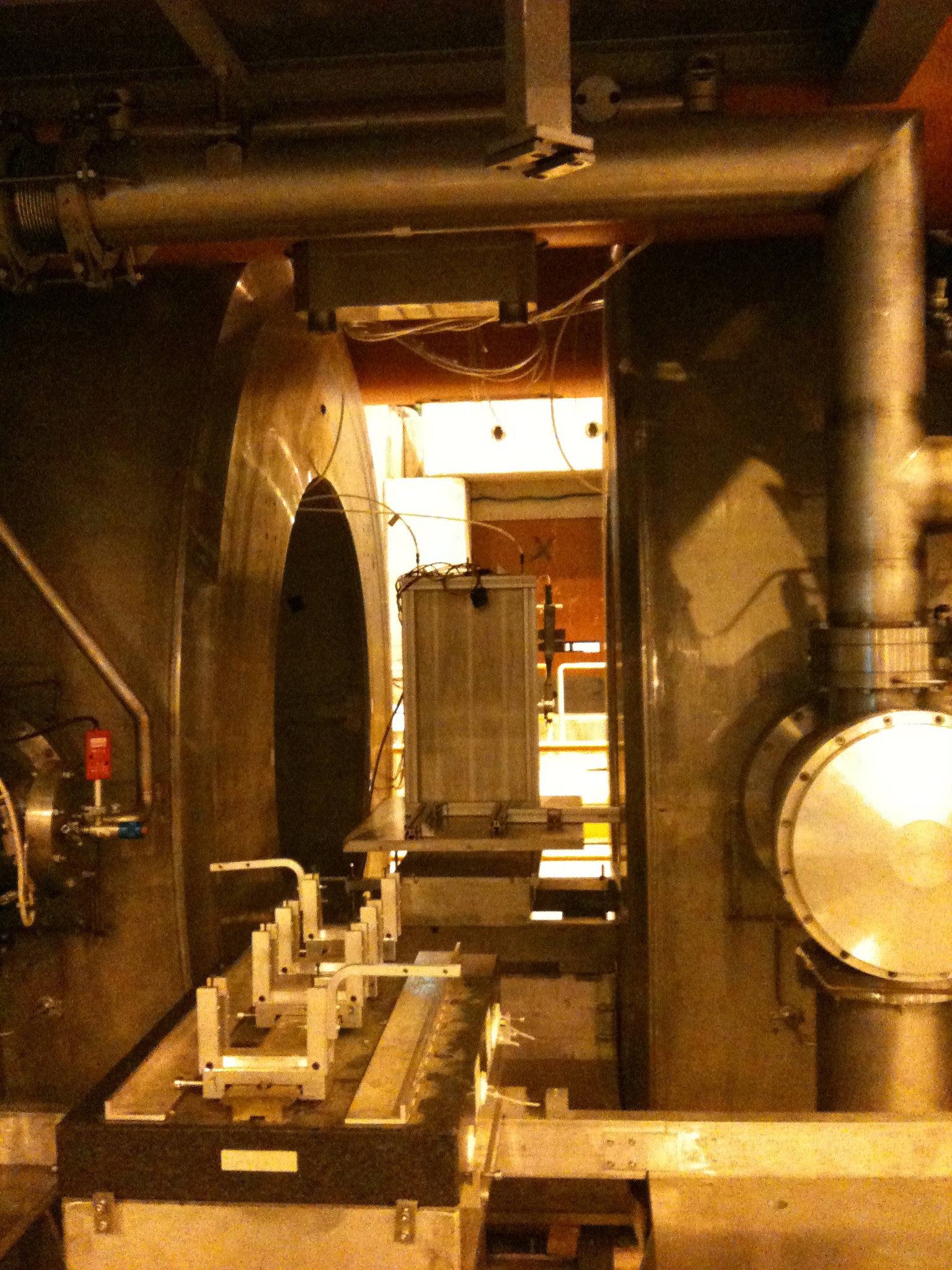}
\end{center}
\caption{The GRPC between the 3-Tesla Helmholtz coils at the CERN SPS H2 beam line.\label{fig.beamline}}
\end{figure}

The PCB of 24 ASICs described in section 2 was attached to the  GRPC detector in such a way that  the pick-out pads of the PCB are in direct contact with the latter. To guarantee good contact  everywhere both the detector and the electronics board were placed inside an aluminum  cassette. The cassette was conceived so that its height corresponds exactly to that of the detector and the electronics board.%
The Detector InterFace board which hosts the FPGA device was connected to the PCB directly and mechanically fixed on the aluminum cassette.   The cassette was then placed in the H2 SPS-CERN beam line  between  the two coils  forming the  H2 super conductucting  magnet as shown in Figure \ref{fig.beamline}.  

\begin{figure}[ht]
      \begin{center}
        \includegraphics*[width=0.6\textwidth,keepaspectratio]{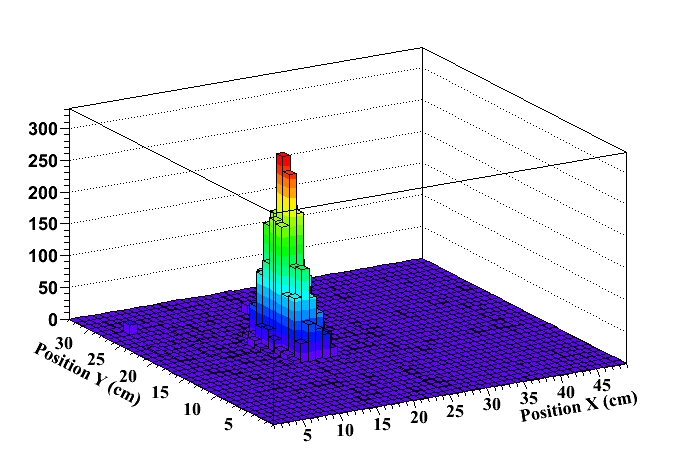}
	\caption{The beam profile as seen by the GRPC detector. \label{beamprofile}}
      \end{center}
\end{figure}

The field produced by this magnet is quite uniform between the two coils (Helmholtz scheme) and it is perpendicular to the beam that passes between the two coils.  The maximum field  that can be reached  by the H2 magnet is  3\u{Tesla} which is very close to the one proposed in the ILD LOI. The orientation of the cassette was chosen to maximize the Lorentz force exercised by the magnetic field on the electronics board. %
A trigger system made of four scintillator-PMs was used. Two of  the scintillator-PMs were placed in front of  the GRPC chamber and two behind with respect to the beam. The  overlap between the four scintillators was chosen so that particles crossing the four scintillators  cross necessarily the GRPC as can be seen in Figure \ref{beamprofile}.  Due to the low detection rate capability of the GRPCs ($<100~\hbox{Hz/cm}^2$)  the beam optics were tuned to reduce the beam profile density to less than 50  particles$/\hbox{cm}^2$.%

To study the behavior of the  power-pulsed electronics on the GRPC detector in a magnetic field, the DAQ was adapted to take into account the time needed for the ASIC modules to be stabilized. The StartAcquisition command  was delayed by 100 $\mu s$ after the POD command was sent. This is twice the delay time mentioned previously in the case of a 24-ASIC, with the  aim of protecting against probable additional perturbations that could be caused by the presence of the strong magnetic field.  In addition, the duty cycle was modified with respect to the ILC duty cycle: 2 ms every 10 ms  rather than every 200 ms. This was intended to amplify the physical effect on the electronics board  due to the repeated increase and decrease of the electric currents circulating in the board. 
Furthermore, to be able to compare the GRPCs efficiency  with and without the power-pulsing mode, the external triggers were limited to the time intervals for which the power-pulsing was on.

\section{Results}
The GRPC equipped with the readout electronics was tested with a pion beam of 150 GeV  first without the power-pulsing and in the absence of a magnetic field.  This allows one to establish a reference behavior of the GRPC.
A gas mixture of tetrafluoroethane (TFE, 93\u{\%}),  $\mathrm{CO_2}$ (5\u{\%}) and $\mathrm{SF_6}$ (2\u{\%}) was used to operate the GRPC.  A common triggering threshold of 140 DAC units was
used for all of the 24 ASICs.   This is equivalent to a 40 fC threshold since the pedestals average was found to be around 100 DACs and the conversion ratio was measured to be around one. With an average value of  $\approx$1 pC/mip and charge spectrum almost negligible up to 100 fC for a polarization voltage of 7.5 kV, this choice of threshold allows to optimize the detector efficiency while keeping the noise rate low.
To estimate the efficiency of  our chamber, the acquisition is stopped upon the arrival of an external trigger. After the readout is performed the recorded hits are associated with their coordinates and their time occurrence determined. If at least one hit is found in a time interval of  2.1$\mu$s preceding the trigger occurrence the hit is associated to the particle at the origin of the trigger. The noise rate was measured to be less than1 Hz/$\hbox{cm}^2$. The signal contamination is thus negligible.    In the absence of a tracking device  to determine the exact impact position of this particle we limit the detector's acceptance zone to the one determined by the overlapping of the four scintillators. 
This allows to get rid of the possible contribution to the efficiency of stray particles crossing the detector in the same window of time associated to the trigger, even though this is improbable due to the low beam intensity used in this test.%

\begin{figure}[ht]
      \begin{center}
        \includegraphics*[width=0.6\textwidth,keepaspectratio]{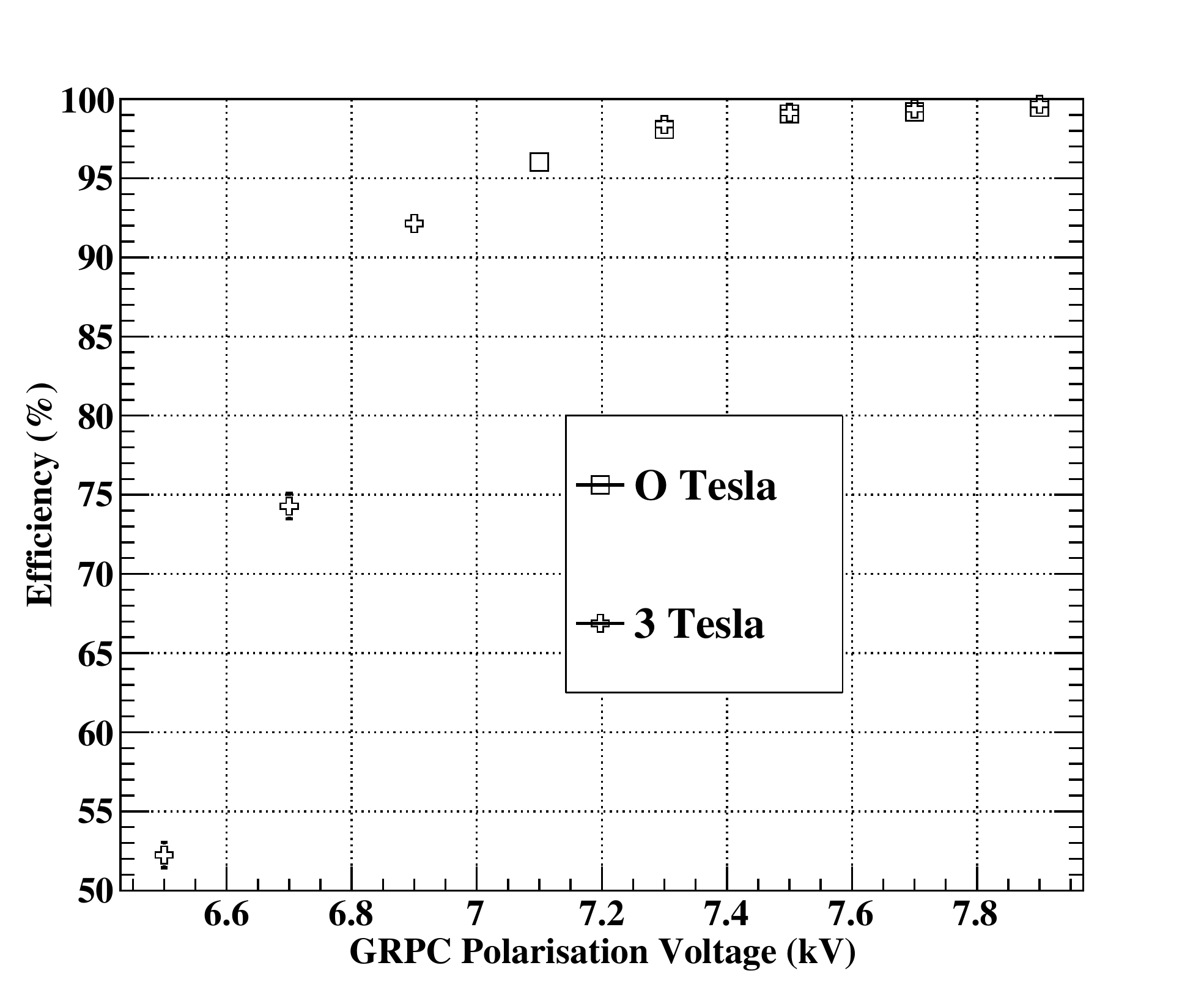}
	\caption{Efficiency scan over high voltage, with and without magnetic field. \label{fig.EffHV1}}
      \end{center}
\end{figure}

The detector efficiency was estimated at different polarization voltage as shown in  Figure \ref{fig.EffHV1}. The shape is similar to that obtained with GRPCs having the same structure but with a smaller size\cite{small}.  At 7.5\u{kV}, the efficiency is about $98~\u{\%}$. %

\begin{figure}[ht]
    \begin{minipage}[t]{0.49\textwidth}
      \begin{center}
        \includegraphics*[width=\textwidth,keepaspectratio]{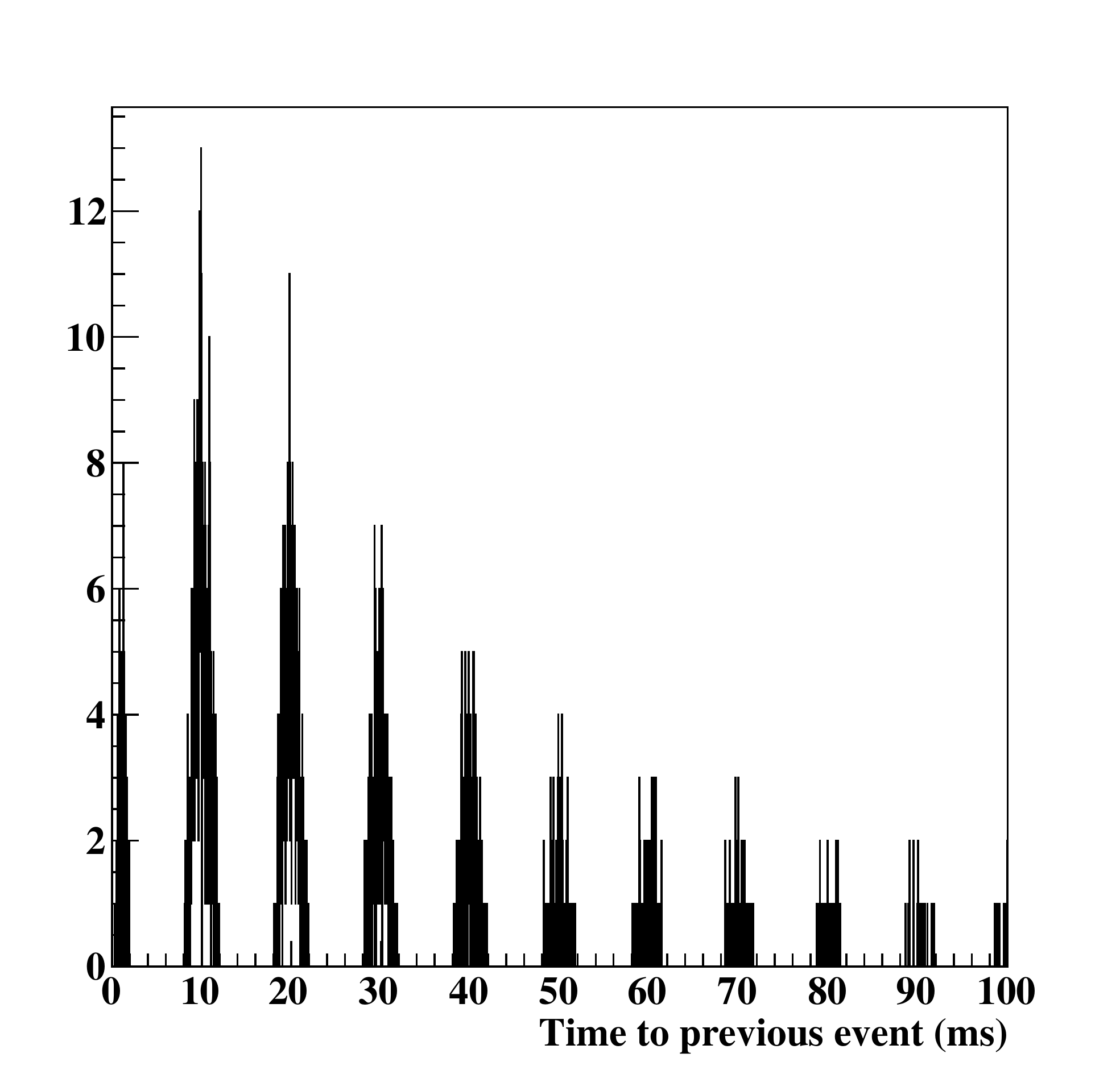}
	\caption{Time difference between consecutive triggers. \label{fig.TimeStruct}} 
      \end{center}
    \end{minipage}
    \hspace{0.02\textwidth}
    \begin{minipage}[t]{0.49\textwidth}
      \begin{center}
        \includegraphics*[width=\textwidth,keepaspectratio]{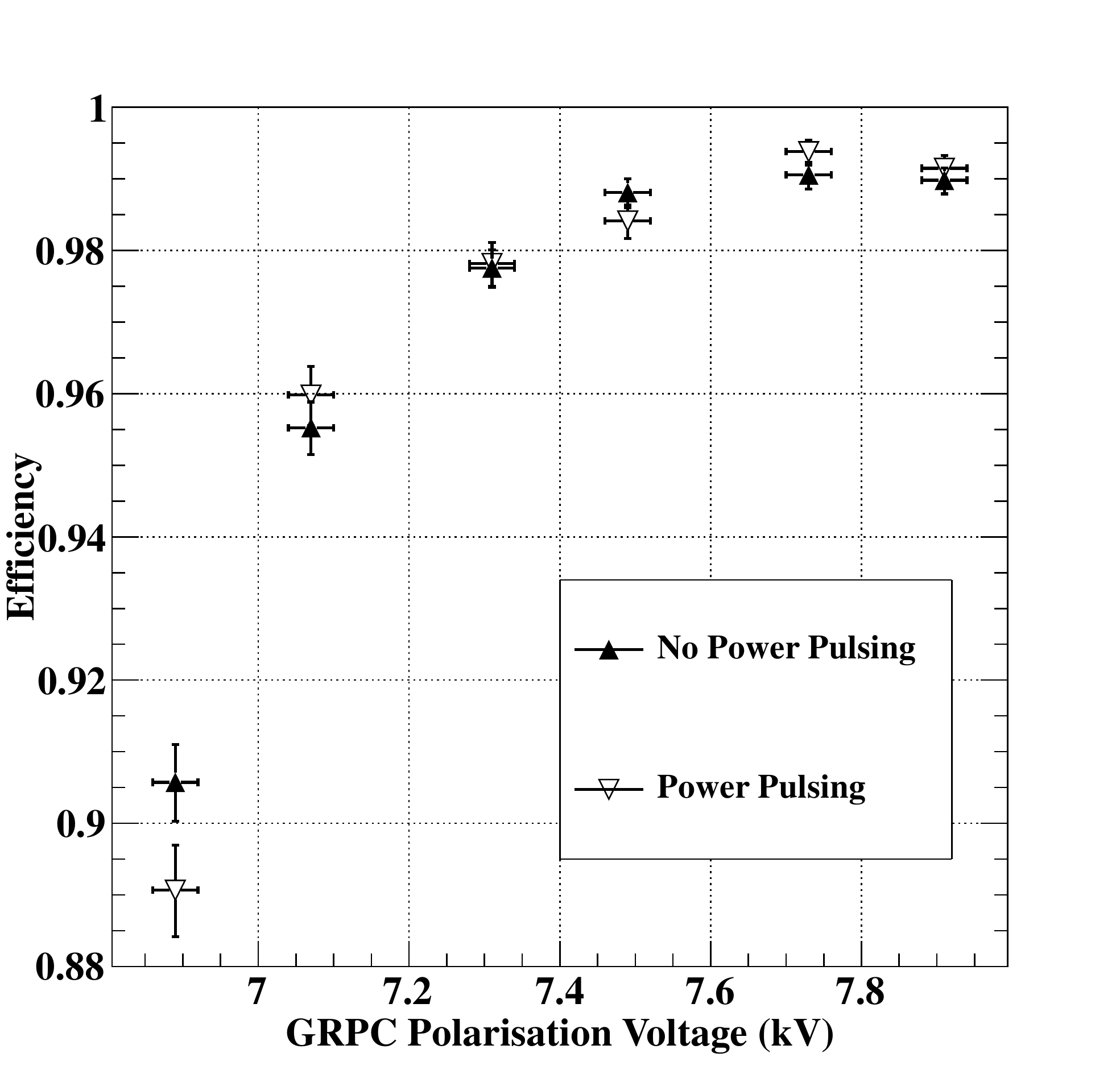}
       \caption{Efficiency scan over high voltage, with and without  power pulsing \label{fig.EffPP}}
      \end{center}
    \end{minipage}
\end{figure}

 The same study was performed again but with the presence of the magnetic field of 3\u{Tesla}. The results in Figure \ref{fig.EffHV1} show clearly that the magnetic field has no effect on the GRPC efficiency, at least for the reference polarization voltage values between 7.3 and 7.8 kV confirming a previous result obtained with similar GRPC detectors operating in a 5\u{Tesla} magnetic field~\cite{Ammosov}. 
 The final comparison was made using the power-pulsing mode with the duty cycle described above in the presence of the magnetic field.  Figure  \ref{fig.TimeStruct}  shows the time structure of the data recorded in the ASICs by taking the time difference of two consecutive triggers.  This shows clearly that the duty cycle mentioned above was respected.  The first peak of the time distribution (Figure \ref{fig.TimeStruct}) corresponds to two events being recorded in the same power cycle, while the second one represents events being recorded in two consecutive power cycles. Figure  \ref{fig.EffPP} shows the efficiency obtained for a polarization voltage range from 6.9 to 8 kV.  The efficiency obtained  using the the PP scheme is almost identical to the one obtained in the absence of this scheme.
 The tiny loss  of efficiency  observed for the 6.9 kV polarization value has no effect  on the performance of the GRPC to be operated around  7.5 kV and thus on the SDHCAL performance. It is however necessary to investigate this in more detail in the near future in order to see if this loss is related to a slight increase of the threshold reference of the comparators after the Power-On-DAC is established, which lasts beyond the 100 $\mu$s protection delay. Indeed the charge associated to a MIP for this polarization (6.9 kV) is slightly lower than the one obtained for the reference value of 7.5 kV and thus an increase of DAC threshold references due to the PP  with respect to the assigned value of  40 fC could affect the efficiency. %

  It is important here to mention that these results show that the variation of  magnetic forces to which the ASUs are subjected  has little effect not only on the ASICs performance but also on the electric and physical proprieties of the electronic boards.  Testing such a board in a magnetic field of 3 T could be detrimental to the overall planarity of the PCB as well as to the board's  different  electrical components. A conductor of length $L$, carrying a current $I$, is submitted to a Lorentz force having modulo $F=B.I.L$ in a magnetic field of $B$ Tesla perpendicular to the current flow: it should be noted then as the dimensions of the board ($50\,\hbox{cm}$ length), in presence of relatively large DC currents, may lead to a force as high as one Newton when B=3\u{T} and I=650\u{mA}.

\section{Conclusions}

A first  demonstration of the possibility of using the power-pulsing mode in the same conditions as those expected in the future ILC experiments was achieved. The results, obtained adopting a duty cycle higher than what foreseen in ILC, show that the power-pulsing mode of the ASIC is operational and does not affect the performance of the future SDHCAL.  A power consumption reduction factor of 100 is then achievable which paves the way for ultra-granularity calorimeter use in the future.  Future test beams using the H2 SPS magnet  are foreseen. Large GRPC detectors similar to the ones to be used for the SDHCAL prototype ~\cite{imad} will be tested using the PP scheme.

\end{document}